\documentclass[twocolumn,showpacs,amsmath,amssymb]{revtex4}
 
\usepackage{graphicx}
\usepackage{dcolumn}
\usepackage{bm}

\begin{document}
\title{Power-Law Time Distribution of Large Earthquakes} 

\author{Mirko S. Mega$^{1}$, Paolo Allegrini$^2$, 
Paolo Grigolini$^{1,3,4}$, Vito Latora$^{5}$\footnote{Corresponding author:
vito.latora@ct.infn.it}, 
Luigi Palatella$^{1}$, Andrea Rapisarda$^{5}$ and Sergio
Vinciguerra$^{5,6}$}

\affiliation{\vspace{.3 cm}$^{1}$Dipartimento di Fisica 
dell'Universit\`a di Pisa and INFM, via Buonarroti 2, 56127 Pisa, Italy}
\affiliation{$^2$Istituto di Linguistica Computazionale del CNR, 
Area della Ricerca di Pisa,  
Via G. Moruzzi 1, 56124, Pisa, Italy}
\affiliation{$^{3}$Center for Nonlinear Science, University of North
Texas,   P.O. Box 311427, Denton, Texas 76203-1427 }
\affiliation{$^4$Istituto di Biofisica del CNR, Area della Ricerca di
Pisa, Via Alfieri 1, San Cataldo,
56010, Ghezzano-Pisa, Italy}
\affiliation{$^{5}$ Dipartimento di Fisica e Astronomia, 
Universit\`a di Catania, 
\\
and INFN sezione di Catania, 
Via S. Sofia 64, 95123 Catania, Italy}
\affiliation{$^{6}$ Osservatorio Vesuviano - INGV, Via Diocleziano 328, 
80124 Napoli, Italy}

\begin{abstract}

We study the statistical properties of time distribution of seimicity  
in California by means of a new method of analysis, the Diffusion Entropy. 
We find that the distribution of time intervals between 
a large earthquake (the main shock of a given seismic sequence)  
and the next one does not obey Poisson statistics, 
as assumed by the current models. 
We prove that this distribution is an inverse power law with an exponent  
$\mu=2.06 \pm 0.01$. We propose the Long-Range model, reproducing the 
main properties of the diffusion entropy and describing the seismic
triggering 
mechanisms induced by large earthquakes.  

\end{abstract}
\maketitle

The search for correlation in the space-time distribution of earthquakes
is a major goal in geophysics. At the short-time and the short-space 
scale the existence of correlation is well established. 
Recent geophysical observations indicate that main fracture episodes  
can trigger long-range as well as short-range seismic effects 
\cite{kagan,hill,cresce,parsons}. 
However, a clear evidence in support of these geophysical indications 
has not yet been provided.  
This is probably the reason why one of the models  
adopted to describe the time distribution of earthquakes 
is still the Generalized Poisson (GP) 
model \cite{shlien,gardner,gasperini,console,godano}. 
Basically the GP model assumes that the earthquakes are grouped into  
temporal clusters of events  
and these {\it clusters are uncorrelated}: 
in fact the clusters are distributed at random in time 
and therefore the time intervals between one cluster and the next one 
follow a Poisson distribution.  
On the other hand, the {\it intra-cluster earthquakes are correlated} 
in time as it is expressed by the Omori's law \cite{omori,utsu}, 
an empirical law stating that the main shock, i.e. the highest 
magnitude earthquake of the cluster, occurring at time $t_{0}$ 
is followed by a swarm of correlated
earthquakes (after shocks) whose number  (or frequency) $n(t)$ decays in
time as a power law, $n(t) \propto (t-t_{0})^{-p}$, with the exponent $p$
being very close to $1$. 
The Omori's law implies \cite{bak} that the distribution of the time
intervals between one earthquake and the next, denoted by $\tau$, is a 
power law $\psi(\tau) \propto \tau^{-p}$. 
This property has been recently
studied by the authors of Ref. \cite{bak} by means of a unified scaling
law
for $\psi_{L,M}(\tau)$, the probability of having a time interval 
$\tau$ between two seismic events with a magnitude larger than $M$
and occurring within a spatial distance $L$. 
This has the effect of taking into account also space and 
extending the correlation within a finite time range
$\tau^{*}$, beyond which the authors of Ref. \cite{bak} 
recover Poisson statistics. 

\begin{figure}
\includegraphics[width=7. cm]{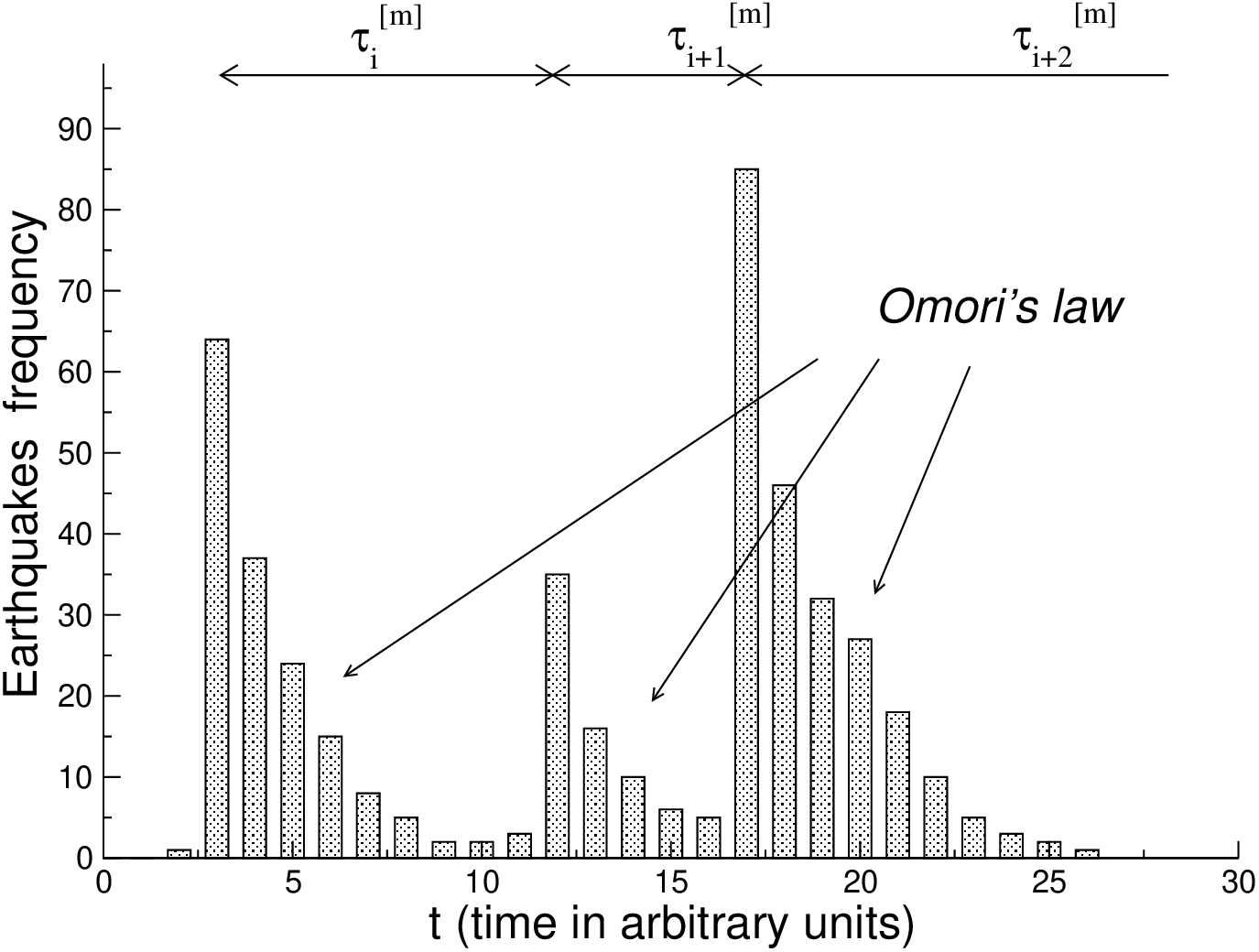}   
\caption{\label{fig1} 
A sketch of the typical earthquakes frequency vs time. 
In  correspondence to each main shock we observe a frequency 
peak determined by the after shock swarm. 
The peaks decay according to the Omori's law. 
The horizontal arrows indicate the time intervals $\tau_i^{[m]}$ between
two 
consecutive main shocks. The DE method gives information on the 
distribution of these time intervals.}
\end{figure}

In this letter, we provide evidence of  
{\it inter-clusters correlation} by studing a catalog of seismic events 
in California with a  new technique of analysis called 
Diffusion Entropy (DE) \cite{scafetta,giacomo}. 
This technique, scarcely sensitive to predictable events such as 
the Omori cascade of aftershocks, is instead very sensitive when the 
deviation from  Poisson statistics  generates L\'{e}vy 
diffusion \cite{giacomo,memory}. This deviation, on the other hand, 
implies that the geophysical process generating clusters   
has some memory. 
In Fig. \ref{fig1} we report the sketch 
of the typical earthquakes frequency vs time in the catalog considered. 
By $\tau_{i}=t_{i+1} - t_{i}$ 
we indicate the time interval between an earthquake and the next.  
Each peak of frequency (cluster) in figure includes the time location 
of a main shock. 
The time interval between one peak and the next is 
reported in figure and is denoted by the symbol 
$\tau_{i}^{[m]}$, where the superscript $m$ stands for main shock.  
We assume that two distinct times $\tau^{[m]}$ are 
not correlated, i.e.  
$\langle \tau_{i}^{[m]} \tau_{j}^{[m]} \rangle  = 
\left \langle \left ( \tau^{[m]} \right )^2 \right \rangle \delta_{i,j}$. 
This assumption does not conflict with the departure 
of the distribution of the times $\tau^{[m]}$  from Poisson. On the other
hand the intercurrence times $\tau_{i}$'s are correlated. 
In fact, after a main shock the earthquake frequency is high. 
Consequently, the $\tau$'s are short and a short value of $\tau$ 
is followed with a large probability by another short value. 
For the same reason we expect that, far from a main shock, 
and prior to the next one, a long value of $\tau$ is followed 
by another long value of $\tau$. 
This implies that the correlation function  
$\langle \tau_{i} \tau_{j} \rangle$ does not decay to zero after 
one step, and that it survives for the whole time interval 
between two consecutive main shocks. 
This means that the long-time relaxation of the correlation
function $\langle \tau_{i} \tau_{j} \rangle$ is determined by the
$\tau^{[m]}$-statistics \cite{memory}, being faster or slower, 
according to whether $\psi(\tau^{[m]})$ is
Poisson or not.  As we shall see, the non-Poissonian condition 
is straightforwardly assessed by the DE, if  $\psi(\tau^{[m]})$ has 
an infinite  second moment. It is important to stress that the 
model of Fig. 1 will be used to support the results of the paper 
with the study of artificial sequences, but in no way it 
implies an \emph{a priori} identification of the main shocks 
for the DE method to work.

The DE method, as almost all the techniques used to detect
correlation in a time series $\xi(t)$, 
is based on the  diffusion
process  of the auxiliary $x$-space through the equation: 
\begin{equation}
\frac{dx}{dt} = \xi(t) .
\label{eom}
\end{equation}      
In the case under study here, the stochastic variable $\xi(t)$ 
is constructed by setting $\xi(t)= 1$ 
(or $\xi(t) = M$, with $M$ being the earthquake magnitude)  
if an earthquake occurs at time $t$,
and setting $\xi(t) = 0$ otherwise. In practice $\xi(t)$ is a string of
long patches of $0$'s occasionally interrupted by $1$'s. With this
prescription we build up a diffusion process in the $x$-space
\cite{stanley}. 
We construct many distinct trajectories, labeled by the integer index $n=
1,2,.$, according to the prescription
  \begin{equation}
     x_n(t)= \int_{n \Delta t}^{n \Delta t + t}
\xi(t^{\prime}) dt^{\prime}, 
     \label{traj}
    \end{equation}                                                                                 
where the generic walker $x_{n}(t)$  takes a step ahead, of either length
$1$ or $M$, every time an earthquake occurs. Note that $\Delta t$ is the
resolution time, set in our case to be equal to $1$ min.  
All the trajectories $x_{n}(t)$ 
occupy at time $t = 0$ the position $x = 0$ and then
spread up over the $x$-axis as a result of their partial or total random
nature.  We study the probability distribution $p(x,t) dx$
of finding the walker position at time $t$ in the interval $[x, x+
dx]$. In
the reference frame moving with velocity $v = W$, where $W$ is the
average number of earthquakes in the time interval $\Delta t$, the
diffusion process is expected to fulfill the scaling condition:
\begin{equation}
   p(x,t) = \frac{1}{t^{\delta}} \, F\left( \frac{x-W
t}{t^{\delta}}\right),
\label{scaling} 
\end{equation}
where $F(y)$ is a function with a form dictated by the statistics of the
process, and $\delta$ is the so-called scaling parameter. 
According to the results of Ref.\cite{scafetta,note} the evaluation of the
scaling parameter $ \delta$ requires the use of  
$S(t) = - \int_{-\infty}^{+\infty} dx ~p(x,t) \ln [p(x,t)]$. 
In fact, using Eq. (\ref{scaling}), we get after 
some simple algebra:    
\begin{equation}
   S(t) = A + \delta  ~ \ln (t).
   \label{keyrelation}
\end{equation}                  
This means that the entropy of the diffusion process is a linear 
function of $\ln (t)$ and a measure of the slope is equivalent to the 
determination of the scaling parameter $\delta$. 
It is now important to observe that the DE method has the 
interesting property of detecting the statistics of really 
random events, as recently discovered by the authors 
of Ref. \cite{memory,lncs}. We refer the reader to these papers for 
 mathematical proofs. 
For the purposes of this letter, the following remarks should suffice. 
The time intervals $\tau$'s are correlated, as shown with the
help of Fig. \ref{fig1}, while the DE method rests on the Shannon
entropy and the Shannon entropy increases only as a consequence of the
occurrence of really random events. 
For the time being we rule out the possibility that the Shannon 
entropy increase is determined by a deterministic bias  
\cite{massi}, and so by a non-stationary condition. 
In the stationary condition, 
the only source of entropy increase is 
given by the occurrence of clusters of seismic events because 
the $\tau^{[m]}$'s are not correlated. 
The deviation from the Poisson statistics are easily detected by 
the DE method if $\psi(\tau^{[m]})$ produces anomalous diffusion. 
Let us consider the non-Poissonian waiting time distribution:
\begin{equation}
\label{inverse}
  \psi(\tau^{[m]}) \approx \frac{1}{({\tau^{[m]})}^{\mu}}  .
  \end{equation}       
The condition $\mu > 3$ implies a finite second moment, 
and consequently ordinary 
diffusion with $\delta = 0.5$. 
The condition $2 \le \mu < 3$, on the contrary, 
produces an infinite second moment, 
and, consequently, through the generalized central 
limit theorem \cite{giacomo}, the anomalous scaling 
\begin{equation}
\label{correctscaling}
\delta=\frac{1}{\mu-1}. 
\end{equation} 
The condition  $1 \le \mu \le 2$ produces an anomalous scaling with 
\begin{equation}
\delta={\mu-1} 
\label{scaling1}
\end{equation}
and would imply non-stationarity (as in  the presence of a 
deterministic bias).   
As we shall see, the non-Poissonian statistics of the distance between 
two clusters (i.e. between large events) is 
detected by the DE method, yielding  the anomalous scaling 
parameter $\delta = 0.94 \pm 0.01$.

The catalog we have studied covers the period 1976-2002 
in the region of Southern California 
spanning $20^0$ N -$45^0$ N latitude and 100$^{0}$ W 125$^{0}$ W 
longitude \cite{scsn}. 
This region is crossed by the most seismogenetic part of the San Andrea 
fault, which accommodates by displacement the primarily strike-slip 
motion between the North America 
and the Pacific plates, producing velocities up to 
$47 mm/yr$ \cite{turcotte}.  
The total number of recorded earthquakes in the catalog is  
$383687$ and includes the June 28 1992 
Landers earthquakes (M = 7.3), the January 17 1994 Northrifge earhquake
(M = 6.7), and the October 16 1999 Hector Mine earthquake (M = 7.1). 
Geophysical observations point out that these large earthquakes 
have triggered a widespread increase of seismic events 
at remote distances in space and in time \cite{hill,parsons}. 
The coupling of the sources of stress change 
(i.e. large earthquakes occurrence) and seismicity triggering mechanisms 
is a primary target of 
geophysical investigations, and, as shown below, is revealed by 
the DE analysis. 


\begin{figure}
\includegraphics[angle=0,width=7. cm]{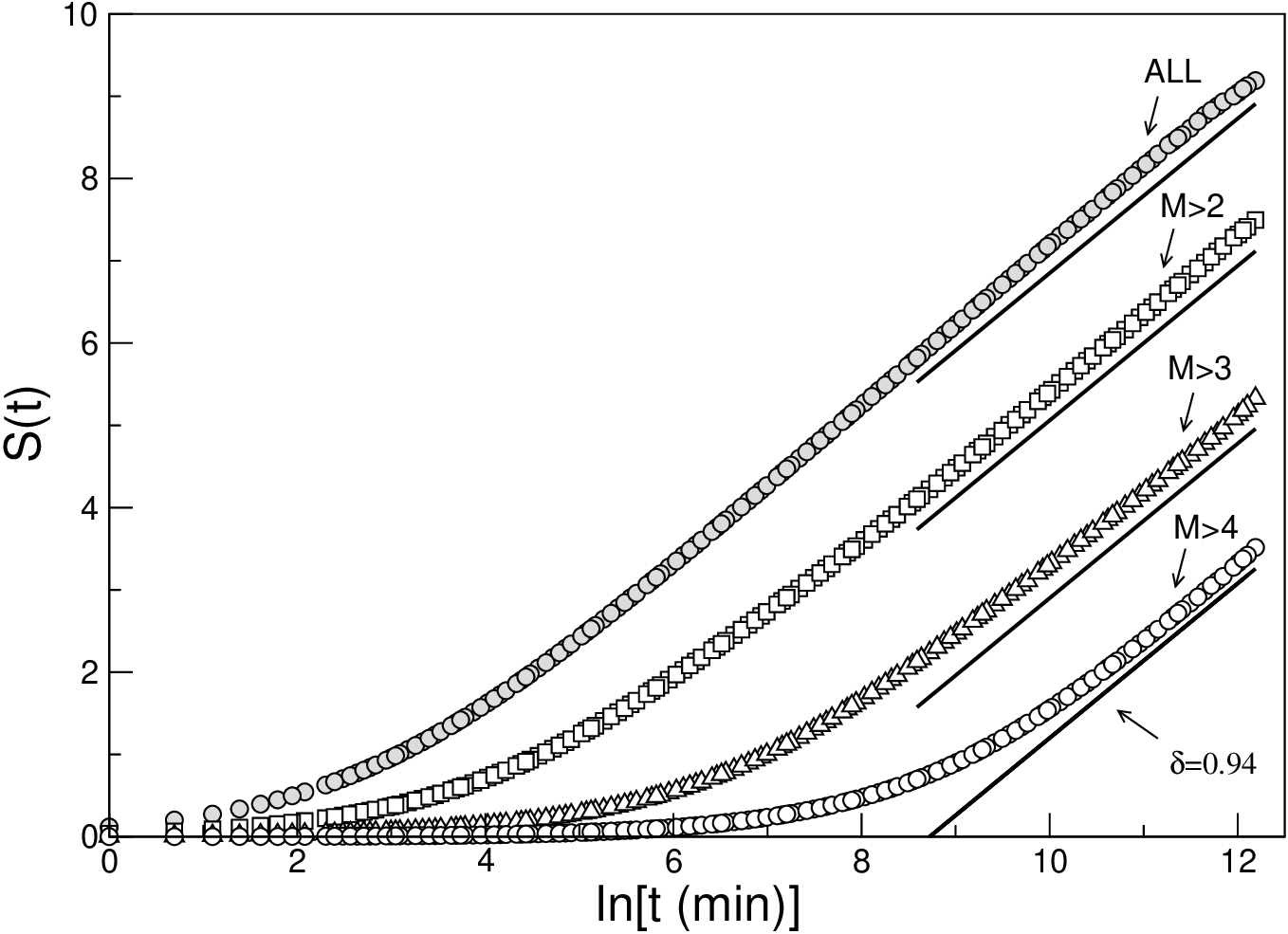}
\caption{\label{fig2} 
The Shannon entropy $S(t)$ of the diffusion process as a function of
time, in a logarithmic time scale. From top to bottom, the curve 
refers to all events (full circles) and to events with 
threshold $\bar M = 2,3,4$ (open symbols). 
The straight lines have the slope $\delta = 0.94$.}
\end{figure}


In Fig. \ref{fig2} we report the results of the DE method. 
The analysis was performed by setting $\xi(t) = 1$ 
when an earthquake occurs at time $t$ 
(independently whether it is a main or an after shock), 
and $\xi(t)=0$ if no earthquake happens. 
In full circles we plot the entropy $S(t)$ as a function of time 
when all the seismic events of the catalog are considered  
(independently of their magnitude $M$). 
After a short transient, the function $S(t)$ is characterized 
by a linear dependence on $\ln t$. 
A fit in the linear region gives a value of the scaling parameter 
$\delta = 0.94 \pm 0.01$ at $95 \%$ of confidence level.
We next consider (open symbols in Fig. \ref{fig2}) only 
the earthquakes with magnitude larger than a fixed value 
$\bar M = 2, 3, 4$. 
We see that, regardless of the value of the threshold $\bar M$
adopted, the function $S(t)$ is characterized by the same long-time
behavior with the same slope.  
This indicates that we are observing 
a property of the time location of large earthquakes. 
This leads us to conclude that the time intervals 
between two large events fit the distribution of Eq.(\ref{inverse}), 
with the value of $\mu$ related to $\delta$ through
Eq.(\ref{correctscaling}), 
$\mu = 2.06 \pm 0.01$. In fact numerical checks on the time series 
under study have supported the 
stationary assumption and  ruled out 
the alternative condition of Eq.(\ref{scaling1}) \cite{note2}.  
Our conclusion is also supported by the use of   
two different walking prescriptions. 
The former rests on assuming $\xi(t)$ equal to the magnitude 
$M$ of the earthquake, at each time when an earthquake occurs. 
The latter sets with equal probability 
either $\xi(t) = 1$ or $\xi(t) = -1$ when an earthquake 
occurs \cite{giacomo}. 
Both methods yield  the same exponent $\mu=2.06 \pm 0.01$. 
%

\begin{figure}
\includegraphics[angle=0,width=7. cm]{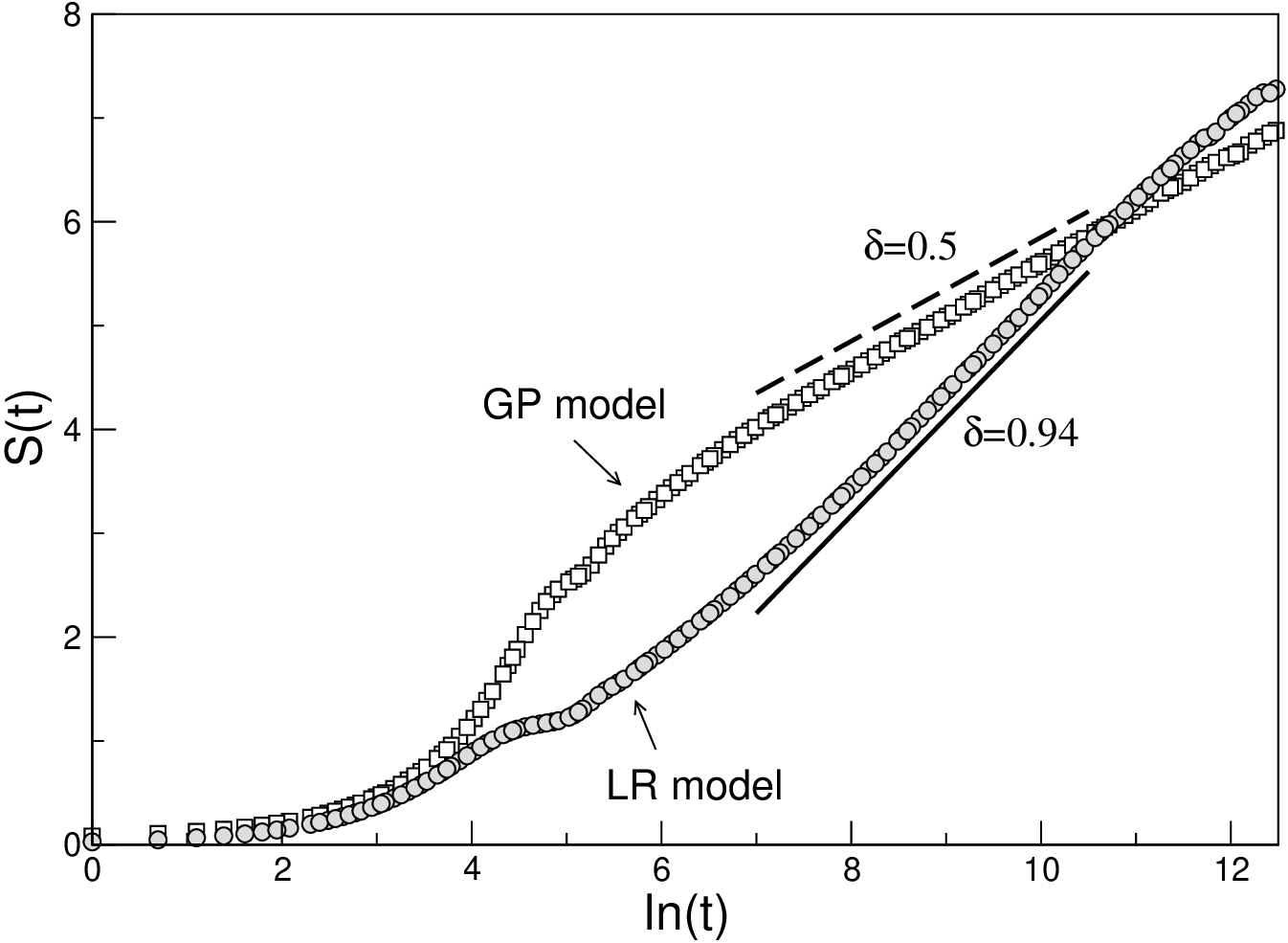}
\caption{\label{fig3} 
The Shannon entropy S(t) of the diffusion process as a function
of time, in a logarithmic time scale. Open squares and full circles 
are, respectively, the results of the GP and of the LR model.   
The two straight lines, have the slopes $\delta =
0.5$ and $\delta = 0.94$.}
\end{figure}
%
The results of our statistical analysis supports the geophysical arguments
that earthquakes of large magnitude produce strain diffusion. 
Unlike co-seismic deformation, which is practically instantaneous, 
the strain diffusion ensuing an earthquake of large magnitude
produces post-seismic stress changes, generating a remarkable increase of 
rate of seismicity at locations hundred of kilometers
away and over time span up to several years. Consistently  
the distribution of the time intervals between two large earthquakes 
is not a Poisson function. 
Our conclusion is reached under the important assumption that 
the sequence $\{  \tau_i^{[m]} \}$ 
is not affected by any deterministic bias, this being a
possible source of ballistic scaling \cite{massi}. 
If this condition applies clusters occurrence would 
show deterministic trends on the scale of the whole sequence.
This is an attractive possibility that does not seem to be ruled out 
by the current literature on this subject \cite{sornette}. 
However, here, we adopt the explanation that the waiting time distribution 
is given by Eq. (\ref{inverse}) with $\mu > 2$, 
this being the unique consequence of 
the stationary assumption.

We now illustrate how the DE method works on two artificial
earthquakes time series: 
the first generated  by means of the GP model, 
and the second generated by a new model, the Long-Range (LR) model, 
that we propose as a better model to reproduce the properties of 
the catolog considered. 
In the LR model the earthquakes are grouped into temporal clusters, 
and, as in the GP model the number of earthquakes in a cluster 
follows the Pareto law, i.e. a power law distribution with exponent 
equal to 2.5 \cite{shlien,gasperini,console}. 
The events within the same cluster are distributed according to 
the Omori's law: the interval $\tau$ follows a power law 
with exponent $p=1$. 
However, in the LR model the time distance $\tau^{[m]}$ 
between one cluster and the next  follows a 
power law with exponent $\mu=2.06$, 
rather than a Poisson prescription as in the GP. 
Notice that this value of $\mu$ 
is close to the border between stationary and non-stationary
condition \cite{giacomo}. 
The two sequences have the same time length. 
We choose the number of clusters in order to have the same 
total number of earthquakes as in the real data \cite{note1}. 
The result of the DE on the artificial sequences is reported 
in Fig. \ref{fig3}. 
The GP model is characterized by a long-time
behavior that, as expected, fits very well the prescription of ordinary
statistical mechanics, with $\delta = 0.5$. The LR model yields the quite
different scaling $\delta = 0.94$. It is also clear that the LR model
yields a
behavior qualitatively similar to that produced by the real data of Fig. 2
as well as the same scaling parameter $\delta = 0.94$, while the GP fail
reproducing both properties.

In conclusion, this paper is the first application of the DE method to 
study the statistical properties of earthquakes time distribution. 
We have found that there exists a correlation  mechanism
beyond the Omori's law. 
Both intra-cluster swarms and inter-cluster distances obey an 
inverse power law prescription, the former being 
$\psi(\tau) \propto \tau^{-1}$ and the latter 
$\psi(\tau^{[m]})  \propto  (\tau^{[m]})^{-\mu}$ 
with $\mu = 2.06 \pm 0.01$.
We have proposed a new model, the LR model, better than the GP model in 
reproducing real data.
The method proposed is based on the fact that the asymptotic properties of 
diffusion process generated by the seismic events are scarsely sensitive 
to the memory stemming from the Omori's law. 
They are, on the contrary, sensitive to the anomalous statistics 
generated by the non-Poissonian nature of the time distance between 
two consecutive large earthquakes.

PG acknowledges support from ARO, through Grant DAAD19-02-0037.

\vfill

\begin{thebibliography}{99}

\bibitem{kagan} Y.Y. Kagan and D.D. Jackson, 
Geophys. J. Int. 104, 117 (1991). 

\bibitem{hill} D.P. Hill et al., Science 260, 1617 (1993).

\bibitem{cresce} L. Crescentini, A. Amoruso, R. Scarpa, 
Science {286}, 2132 (1999). 

\bibitem{parsons} T. Parsons, J. Geophys. Res., 107, 2199 (2001).

\bibitem{shlien} S. Shlien, M.N. Toksoz, 
Bull. Seism. Soc. Am., 60, 1765 (1970).

\bibitem{gardner} J.K. Gardner and L. Knopoff, 
Bull. Seism. Soc. Am., 64, 1363 (1974). 

\bibitem{gasperini} P. Gasperini, F. Mulargia, 
Bull. Seism. Soc. Am., 79, 973 (1989).

\bibitem{console}
R. Console, M. Murru, J. Geophys. Res., 106, 8699, (2001).

\bibitem{godano}
C. Godano and V. Caruso, Geophys. J. Int. 121, 385 (1995).

\bibitem{omori} 
F. Omori, J. College Sci. Imper. Univ. Tokyo  7, 111 (1895).

\bibitem{utsu}  
T. Utsu, Geophys. Mag., 30, 521 (1961).
shock
\bibitem{bak} P. Bak, K. Christensen, L. Danon and T. Scanlon, 
 Phys. Rev. Lett {88}, 178501 (2002).

\bibitem{scafetta}
N. Scafetta, P. Hamilton,
P. Grigolini, Fractals, 9, 193-208 (2001).

\bibitem{giacomo}
 P. Grigolini, L. Palatella, G.
Raffaelli, Fractals, 9, 439-449 (2001).

\bibitem{memory}
 P. Allegrini, P. Grigolini, P.
Hamilton, L. Palatella, and G. Raffaelli, Phys. Rev. E65, 041926 (2002).

\bibitem{stanley} S.V. Buldyrev, A.L. Goldberger, S. Havlin, 
C.-K. Peng, M. Simons, and H.E. Stanley, 
Phys. Rev. E  47, 4514 (1993).

\bibitem{note} The work of Ref.\cite{scafetta} proves that the 
adoption of the Tsallis entropy would be beneficial to 
explore the transition to the scaling regime, but it 
would not determine the scaling parameter $\delta$. 

\bibitem{lncs}
P. Allegrini, R. Balocchi, S. Chillemi, P. Grigolini, L. Palatella, 
G. Raffaelli, 
LectureNotes Computer Sciences 2526, 115-126 A. Colosimo et al. (Eds.) 
Springer-Verlag Berlin Heidelberg 2002.

\bibitem{massi}  M. Ignaccolo, P. Allegrini, P. Grigolini, P. Hamilton,
B.J.
West, submitted to Physica A

\bibitem{scsn} 
The catalog has been downloaded from 
the Southern California Earthquake Data Centre 
http://www.scecdc.scec.org/ftp/catalogs/SCSN/ 

\bibitem{turcotte} 
D. L. Turcotte and G. Schubert, 
Geodynamics, Cambridge University Press 2002. 

\bibitem{note2}
In any case, by using Eq.(\ref{scaling1}) one would 
obtain an exponent $\mu = 1.94 \pm 0.01 $, a value 
well different from the Omori's law exponent.

\bibitem{sornette}
D. Sornette and A. Helmstetter, Phys. Rev. Lett.  89, 158501
(2002)

\bibitem{note1} 
We made other tests by adopting different exponents of 
the Pareto's law ($3.0 \pm 0.7$) and different 
number of clusters, obtaining different transient behavior but 
the same value of $\delta$.



\end{thebibliography}
\end{document}